\documentclass[twocolumn,twocolappendix]{aastex63}
\usepackage{xcolor}

\newcounter{daggerfootnote}

\newcommand{\jwst}{\textit{JWST}}

\definecolor{mycol}{rgb}{0,0,1}

\received{n/a}
\revised{n/a}
\accepted{n/a}

\submitjournal{\apjl}

\shorttitle{UNCOVER z=8.5 AGN }
\shortauthors{Kokorev et al.}

\usepackage{longtable}
\usepackage{lipsum}
\usepackage{amsmath}
\usepackage[normalem]{ulem}
\usepackage[para]{threeparttable}
\usepackage{enumitem}
\usepackage[encapsulated]{CJK}

\begin{document}

\title{UNCOVER: A NIRSpec Identification of a Broad Line AGN at \boldmath $z=8.50$}

\correspondingauthor{Vasily Kokorev}
\email{kokorev@astro.rug.nl}

\author[0000-0002-5588-9156]{Vasily Kokorev}
\affiliation{Kapteyn Astronomical Institute, University of Groningen, 9700 AV Groningen, The Netherlands}

\author[0000-0001-7201-5066]{Seiji Fujimoto}
\affiliation{Department of Astronomy, The University of Texas at Austin, Austin, TX 78712, USA}
\affiliation{Cosmic Dawn Center (DAWN), Niels Bohr Institute, University of Copenhagen, Jagtvej 128, K{\o}benhavn N, DK-2200, Denmark}

\author[0000-0002-2057-5376]{Ivo Labbe}
\affiliation{Centre for Astrophysics and Supercomputing, Swinburne University of Technology, Melbourne, VIC 3122, Australia}

\author[0000-0002-5612-3427]{Jenny E. Greene}
\affiliation{Department of Astrophysical Sciences, Princeton University, 4 Ivy Lane, Princeton, NJ 08544}

\author[0000-0001-5063-8254]{Rachel Bezanson}
\affiliation{Department of Physics and Astronomy and PITT PACC, University of Pittsburgh, Pittsburgh, PA 15260, USA}

\author[0000-0001-8460-1564]{Pratika Dayal}
\affiliation{Kapteyn Astronomical Institute, University of Groningen, 9700 AV Groningen, The Netherlands}

\author[0000-0002-7524-374X]{Erica J. Nelson}
\affiliation{Department for Astrophysical and Planetary Science, University of Colorado, Boulder, CO 80309, USA}


\author[0000-0002-7570-0824]{Hakim Atek}
\affiliation{Institut d'Astrophysique de Paris, CNRS, Sorbonne Universit\'e, 98bis Boulevard Arago, 75014, Paris, France}

\author[0000-0003-2680-005X]{Gabriel Brammer}
\affiliation{Cosmic Dawn Center (DAWN), Niels Bohr Institute, University of Copenhagen, Jagtvej 128, K{\o}benhavn N, DK-2200, Denmark}

\author[0000-0001-8183-1460]{Karina I. Caputi}
\affiliation{Kapteyn Astronomical Institute, University of Groningen, 9700 AV Groningen, The Netherlands}
\affiliation{Cosmic Dawn Center (DAWN), Niels Bohr Institute, University of Copenhagen, Jagtvej 128, K{\o}benhavn N, DK-2200, Denmark}

\author[0009-0009-9795-6167]{Iryna Chemerynska}
\affiliation{Institut d'Astrophysique de Paris, CNRS, Sorbonne Universit\'e, 98bis Boulevard Arago, 75014, Paris, France}

\author[0000-0002-7031-2865]{Sam E. Cutler}
\affiliation{Department of Astronomy, University of Massachusetts, Amherst, MA 01003, USA}

\author[0000-0002-1109-1919]{Robert Feldmann}
\affiliation{Institute for Computational Science, University of Zurich, Zurich, CH-8057, Switzerland}

\author[0000-0001-7440-8832]{Yoshinobu Fudamoto} 
\affiliation{Waseda Research Institute for Science and Engineering, Faculty of Science and Engineering, Waseda University, 3-4-1 Okubo, Shinjuku, Tokyo 169-8555, Japan}
\affiliation{National Astronomical Observatory of Japan, 2-21-1, Osawa, Mitaka, Tokyo, Japan}

\author[0000-0001-6278-032X]{Lukas J. Furtak}
\affiliation{Physics Department, Ben-Gurion University of the Negev, P.O. Box 653, Be'er-Sheva 84105, Israel}

\author[0000-0003-4700-663X]{Andy D. Goulding}
\affiliation{Department of Astrophysical Sciences, Princeton University, 4 Ivy Lane, Princeton, NJ 08544}

\author[0000-0002-2380-9801]{Anna de Graaff}
\affiliation{Max-Planck-Institut f{\"u}r Astronomie, K{\"o}nigstuhl 17, D-69117, Heidelberg, Germany}

\author[0000-0001-6755-1315]{Joel Leja}
\affiliation{Department of Astronomy \& Astrophysics, The Pennsylvania State University, University Park, PA 16802, USA}
\affiliation{Institute for Computational \& Data Sciences, The Pennsylvania State University, University Park, PA 16802, USA}
\affiliation{Institute for Gravitation and the Cosmos, The Pennsylvania State University, University Park, PA 16802, USA}

\author[0000-0001-9002-3502]{Danilo Marchesini}
\affiliation{Department of Physics \& Astronomy, Tufts University, MA 02155, USA}

\author[0000-0001-8367-6265]{Tim B. Miller}
\affiliation{Department of Astronomy, Yale University, New Haven, CT 06511, USA}
\affiliation{Center for Interdisciplinary Exploration and Research in Astrophysics (CIERA) and
Department of Physics and Astronomy, Northwestern University, 1800 Sherman Ave, Evanston IL 60201, USA}

\author[0000-0003-2804-0648]{Themiya Nanayakkara}
\affiliation{Centre for Astrophysics and Supercomputing, Swinburne University of Technology, Melbourne, VIC 3122, Australia}

\author[0000-0001-5851-6649]{Pascal A. Oesch}
\affiliation{Department of Astronomy, University of Geneva, Chemin Pegasi 51, 1290 Versoix, Switzerland}
\affiliation{Cosmic Dawn Center (DAWN), Niels Bohr Institute, University of Copenhagen, Jagtvej 128, K{\o}benhavn N, DK-2200, Denmark}

\author[0000-0002-9651-5716]{Richard Pan}
\affiliation{Department of Physics \& Astronomy, Tufts University, MA 02155, USA}

\author[0000-0002-0108-4176]{Sedona H. Price}
\affiliation{Department of Physics and Astronomy and PITT PACC, University of Pittsburgh, Pittsburgh, PA 15260, USA}

\author[0000-0003-4075-7393]{David J. Setton}
\affiliation{Department of Physics and Astronomy and PITT PACC, University of Pittsburgh, Pittsburgh, PA 15260, USA}

\author[0000-0001-8034-7802]{Renske Smit}
\affiliation{Astrophysics Research Institute, Liverpool John Moores University, 146 Brownlow Hill, Liverpool L3 5RF, UK}

\author[0000-0001-7768-5309]{Mauro Stefanon}
\affiliation{Departament d'Astronomia i Astrof\`isica, Universitat de Val\`encia, C. Dr. Moliner 50, E-46100 Burjassot, Val\`encia,  Spain}
\affiliation{Unidad Asociada CSIC "Grupo de Astrof\'isica Extragal\'actica y Cosmolog\'ia" (Instituto de F\'isica de Cantabria - Universitat de Val\`encia)}

\author[0000-0001-9269-5046]{Bingjie Wang (\begin{CJK*}{UTF8}{gbsn}王冰洁\ignorespacesafterend\end{CJK*})}
\affiliation{Department of Astronomy \& Astrophysics, The Pennsylvania State University, University Park, PA 16802, USA}
\affiliation{Institute for Computational \& Data Sciences, The Pennsylvania State University, University Park, PA 16802, USA}
\affiliation{Institute for Gravitation and the Cosmos, The Pennsylvania State University, University Park, PA 16802, USA}

\author[0000-0003-1614-196X]{John R. Weaver}
\affiliation{Department of Astronomy, University of Massachusetts, Amherst, MA 01003, USA}

\author[0000-0001-7160-3632]{Katherine E. Whitaker}
\affiliation{Department of Astronomy, University of Massachusetts, Amherst, MA 01003, USA}
\affiliation{Cosmic Dawn Center (DAWN), Niels Bohr Institute, University of Copenhagen, Jagtvej 128, K{\o}benhavn N, DK-2200, Denmark}

\author[0000-0003-2919-7495]{Christina C. Williams}
\affiliation{NSF’s National Optical-Infrared Astronomy Research Laboratory, 950 N. Cherry Avenue, Tucson, AZ 85719, USA}
\affiliation{Steward Observatory, University of Arizona, 933 North Cherry Avenue, Tucson, AZ 85721, USA}

\author[0000-0002-0350-4488]{Adi Zitrin}
\affiliation{Physics Department, Ben-Gurion University of the Negev, P.O. Box 653, Be'er-Sheva 84105, Israel}

\begin{abstract}
Deep observations with \textit{JWST} have revealed an emerging population of red point-like sources that could provide a link between the postulated supermassive black hole seeds and observed quasars. In this work we present a \textit{JWST}/NIRSpec spectrum from the JWST Cycle 1
UNCOVER Treasury survey, of a massive accreting black hole at $z=8.50$, displaying a clear broad-line component as inferred from the H$\beta$ line with FWHM = $3439\pm413$ km s$^{-1}$, typical of the broad line region of an active galactic nucleus (AGN). The AGN nature of this object is further supported by high ionization, as inferred from emission lines, and a point-source morphology. We compute the black hole mass of log$_{10}(M_{\rm BH}/M_\odot)=8.17\pm0.42$, and a bolometric luminosity of $L_{\rm bol}\sim6.6\times10^{45}$ erg s$^{-1}$. These values imply that our object is accreting at $\sim 40\%$ of the Eddington limit. Detailed modeling of the spectral energy distribution in the optical and near-infrared, together with constraints from ALMA, indicate an upper limit on the stellar mass of log$_{10}(M_{\rm *}/M_\odot)<8.7$, which would lead to an unprecedented ratio of black hole to host mass of at least $\sim 30 \%$. This is orders of magnitude higher compared to the local QSOs, but is consistent with recent AGN studies at high redshift with \jwst.
This finding suggests that a non-negligible fraction of supermassive black holes either started out from massive seeds and/or grew at a super-Eddington rate at high redshift. Given the predicted number densities of high-$z$ faint AGN, future NIRSpec observations of larger samples will allow us to further investigate the galaxy-black hole co-evolution in the early Universe.
\end{abstract}


\section{Introduction} \label{sec:intro}
Over the past decades, observations have established a sample of more than 200 bright active galactic nuclei (AGN) at $z>6$, powered by accretion onto massive black holes \citep[e.g.][]{fan01,kashikawa15,matsuoka18}. Lying well within the first billion years, many of these black holes are massive \citep[$\sim10^{8-10}$ $M_\odot$;][]{banados18,inayoshi20} with the heaviest black hole having a mass of about $1.6\times 10^9$ only 700 million years after the Big Bang \citep{wang21}. The presence of such super massive black holes (SMBHs) is extremely hard to reconcile with black hole formation and growth scenarios since they require extremely massive seeds (of $\sim 1300$ $M_\odot$) to form shortly after the Big Bang and then continuously and rapidly accrete gas at the Eddington rate (a physical limit at which outward radiation pressure balances inward gravitational force). One possible solution to explain the presence of such SMBHs lies in invoking super-Eddington accretion rates \citep[e.g.][]{haiman01,alexander14} to drive the growth of low mass black hole seeds ($\sim 100$ $M_\odot$) formed from the first (metal-free Population III) stars \citep{madau01}. Another possible solution is to start from massive seeds which includes ``nuclear clusters" of $10^{2-4}$ $M_\odot$ from the collapse or coalescence of massive stars in compact stellar clusters \citep{omukai08,schleicher22} or even heavier ``direct collapse black hole seeds" of $>10^3$ M$_{\odot}$ from the collapse of pristine gas in early halos \citep{rees84,loeb94}.
\par
Observations with the \emph{James Webb Space Telescope} (JWST) have started to discover previously missing UV-faint AGN. These have been identified through a combination of broad and high ionization lines \citep{furtak23_nat,goulding23,harikane23_agn,kocevski23,larson23,maiolino23,maiolino23b,ubler23} or from color and morphology \citep{barro23,matthee23,yang23}. In common across these selections are extremely red rest-frame optical colors, flat $f_\lambda$ UV continuum  and generally a very compact morphology. When spectra are available, broad Balmer series lines, such as H$\alpha$ or H$\beta$ are a tell-tale sign of ongoing accretion. While these are UV-faint, they span a large range of bolometric luminosities $L_{\rm bol} \sim 10^{43}-10^{46}$~erg s$^{-1}$ and black hole (BH) masses $M_{\rm BH}\sim 10^6-10^8$~$M_\odot$, with some of them being strongly dust obscured, up to $A_{\rm V}\sim3$ \citep{furtak23_nat}. In addition, recently 
\citet{labbe23_nat,labbe23} presented a large photometrically selected sample of AGN in the UNCOVER Cycle I JWST program. These have been modeled with a composite SED model, which consists of a dust-reddened Type I AGN, plus an additional UV component, attributed to either the scattered light or star formation. The unique finding present across all of these works are the high number densities of reddened QSOs, suggesting than a non-negligible fraction of BH growth at these epochs is taking place behind a thick veil of dust.
\par
In this work we report a secure detection of the broad-line AGN emission at $z=8.50$. The point source hosting the AGN was first identified as one of the ``little red-dots" \citep{greene23,labbe23} in the UNCOVER NIRCam sample. Here we report the results from $JWST$/NIRSpec Micro-Shutter Assembly (MSA) observations of this source in the UNCOVER field (PIs: I. Labb\'e, R. Bezanson; \citealt{bezanson22}).
The exquisite high S/N NIRSpec spectrum allows us to resolve a multitude of lines, including $H\beta$ and [OIII] $\lambda\lambda$ 4959,5007 \AA{}, and clearly demonstrate the presence of the broad line emission in $H\beta$. The [OIII] $\lambda\lambda$ 4959,5007 \AA\, lines however only display narrow emission, which in conjunction with a near point-like morphology and presence of highly ionized gas provide strong evidence for AGN activity in our source. A clear broad line (BL) emission in $H\beta$ would make our source the highest redshift BL AGN with a secure (S/N$>10$) Balmer BL identification to date \citep[see also][]{larson23}.
\par
Throughout this work we assume a flat $\Lambda$CDM cosmology with $\Omega_{\mathrm{m},0}=0.3$, $\Omega_{\mathrm{\Lambda},0}=0.7$ and H$_0=70$ km s$^{-1}$ Mpc$^{-1}$, and a \citet{chabrier} initial mass function (IMF) between $0.1-100$ $M_{\odot}$. All magnitudes are expressed in the AB system \citep{oke74}.

\begin{figure*}
\begin{center}
\includegraphics[width=.95\textwidth]{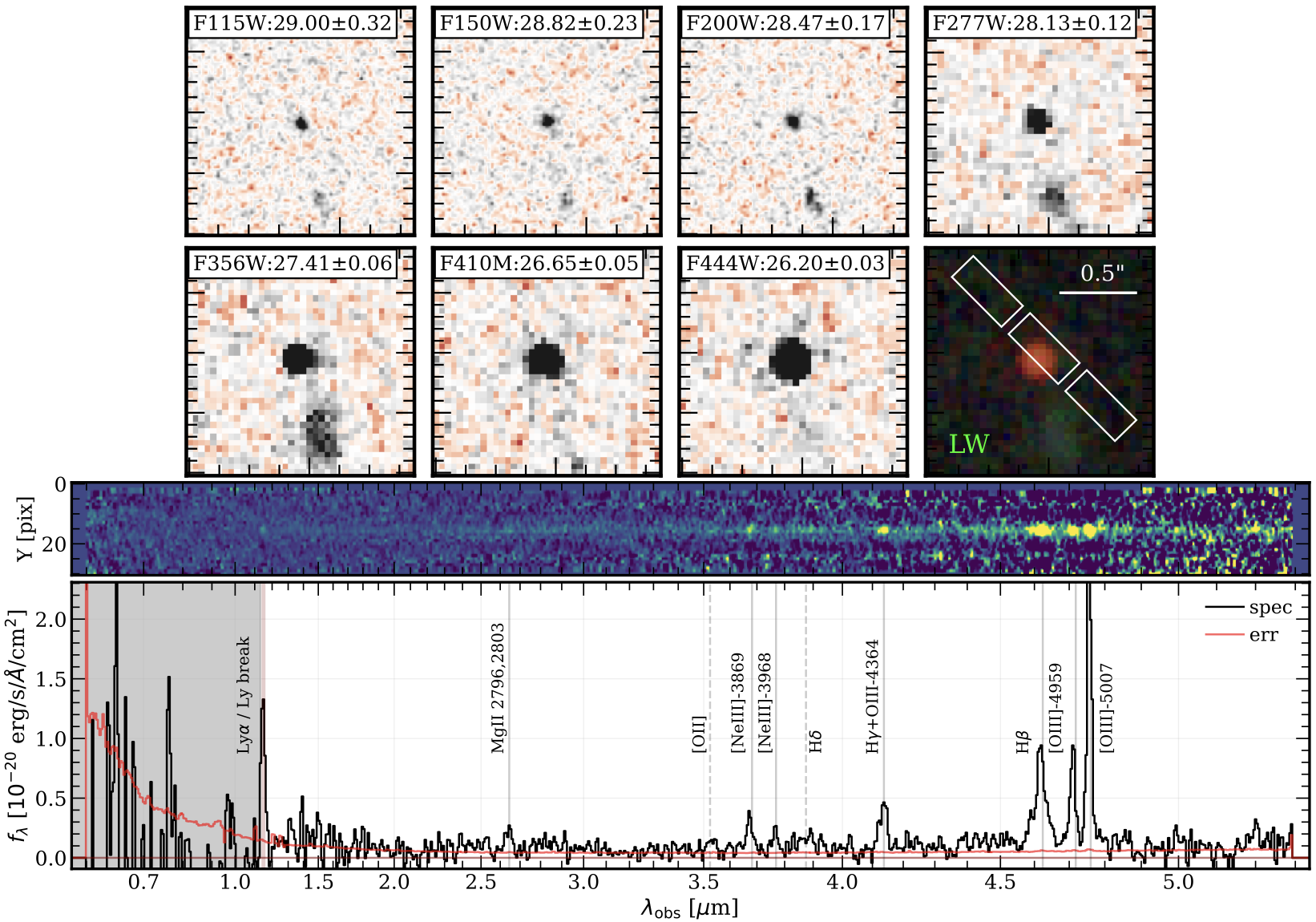}
\caption{\textbf{Top:}
\textit{JWST}/NIRCam 1\farcs{5} stamps and the RGB color image comprised of the F277W, F356W and F444W bands. The MSA slitlet layout is highlighted in white. An unambiguous point-like morphology of ID: 20466 can be observed in all filters. On each panel we show total magnitudes, with $1\sigma$ uncertainty, as presented in the UNCOVER photometric catalog of \citet{weaver23}.
\textbf{Middle:} 2D MSA PRISM spectrum produced by \textsc{msaexp}. We optimally scaled the trace to highlight all the significant line detections.
\textbf{Bottom:} A collapsed 1D spectrum of our source. We show the data in black, while the uncertainty on the spectrum is in red. Assuming the best-fit \textsc{msaexp} $z_{\rm spec}=8.502\pm0.003$ we show the positions of all the prominent emission with significant ($>3\sigma$) detections as solid vertical lines. Emission for which we only obtain an upper limit are shown with dashed lines.}
\label{fig:fig1}
\end{center}
\end{figure*}

\section{Observations and Data} \label{sec:obs_data}
\label{sec:data}

\subsection{NIRSpec Setup}
Initially identified as a potential high-$z$ AGN candidate via the broad and medium band NIRCam photometry \citep{weaver23,labbe23_nat,labbe23}, UNCOVER ID: 20466\footnote{ Corresponds to the MSA ID, taken from internal UNCOVER catalog v2.2.1} is located at R.A. = 3.640408$^{\circ}$, Dec = -30.386438$^{\circ}$. The target was observed for a total of 2.7 hours with the low-resolution PRISM on July 31 2023, as a part of the MSA follow-up program of the UNCOVER JWST field, Abell 2744 \citep{bezanson22}. These observations employed a 2-POINT-WITH-NIRCam-SIZE2 dither pattern and a 3 shutter slitlet nod pattern at an aperture angle of $\sim$44.0$^{\circ}$. We show the positions of these slits overlaid on a 2\farcs{0} cutout of our source in \autoref{fig:fig1}. For further details of the observational setup see \citet{bezanson22} and Price et al. (2023, in prep).

\subsection{PRISM Data Reduction and Calibration}

The UNCOVER micro-shutter array (MSA) spectra were reduced by using the \textsc{msaexp} \citep[v0.6.10,][]{msaexp}, starting from the level 2 data products obtained from MAST\footnote{Available from: \url{http://dx.doi.org/10.17909/8k5c-xr27}}. The pipeline applies corrections to the $1/f$ noise, identifies artifacts (e.g. ``snowballs") in the calibrated files, as well as removes bias in individual exposures by computing a simple median (e.g. see \citealt{rigby23}). Additional parts of the \textit{JWST} reduction pipeline are used to set the slit WCS, perform slit-level flat-fielding and compute the path-loss corrections. Each of the 2D slits are then extracted and drizzled onto a common pixel grid. The local background subtraction is performed by using the shifted, stacked 2D spectra. To obtain the final 1D spectrum, \textsc{msaexp} performs an adaptive optimal extraction of the background subtracted 2D trace by fitting a Gaussian model where the center and width are allowed to vary \citep[e.g.,][]{horne86}, a similar procedure was used in 
\citet{goulding23} and \citet{wang23}.
\par
It is important to note that the absolute normalization of the spectrum can be affected by many factors, which include, but are not necessarily limited to the path-loss correction, calibration uncertainty\footnote{Can be of the order of 10-20\%: \url{https://jwst-docs.stsci.edu/jwst-near-infrared-spectrograph}}, the astrometric slit position uncertainty, the location and extent of the source within a slitlet, and finally source self-subtraction when correcting for the local background effects. To account for these potential calibration issues and to determine a total slitloss correction, we perform an additional flux rescaling by convolving the extracted 1D spectrum with all the broad/medium band NIRCam filters and compare our flux densities to the total photometry presented in \citet{weaver23}. The wavelength dependent linear correction is then computed by fitting a first order polynomial. A complete description of the data reduction and flux calibration will be presented in Price et al. (2023, in prep.).

\begin{deluxetable}{cc}[]
\tabcolsep=4mm
\tablecaption{\label{tab:tab1}
	Source Properties$^\dagger$}.
\tablehead{}
\startdata
R.A. & $3.640408$ \\
Dec & $-30.386438$\\
$z_{\rm spec}$ &  $8.502\pm0.003$\\
$z_{\rm phot}$ $^1$ &  $9.0\pm0.3$\\
$\mu$ & $1.33\pm0.02$\\
$A_{\rm V}$ (H$\beta$/H$\gamma$) [mag] & $2.1^{+1.1}_{-1.0}$ \\
$A_{\rm V}$ (continuum) [mag] & $\sim1.9$ \\
log$_{10}(M_{*}/M_{\odot})$ & $<8.7$ \\
FWHM$_{\rm narrow}$ [km s$^{-1}$]& $203\pm154$ \\
FWHM$_{\rm broad}$ [km s$^{-1}$] & $3439\pm413$ \\
log$_{10}(M_{\rm BH}/M_{\odot})$ (H${\beta}$) & $8.17\pm0.42$ \\
log$_{10}(M_{\rm BH}/M_{\odot})$ ($L_{5100}$) & $8.01\pm0.40$ \\
$L_{\rm bol}$ [erg s$^{-1}$] & $(6.6\pm3.1)\times10^{45}$ \\
Ly$\alpha$ EW$_{\rm 0}$ [\AA] & 240$\pm$30\\
\enddata
\begin{tablenotes}
$^\dagger$ \footnotesize{Corrected for the lensing magnification.} \\
\item[1] \footnotesize{From the latest UNCOVER v3.0.1 catalog}
\end{tablenotes}
\end{deluxetable}

\section{Data Analysis}
\label{sec:analysis}

\begin{figure*}
\begin{center}
\includegraphics[width=.95\textwidth]{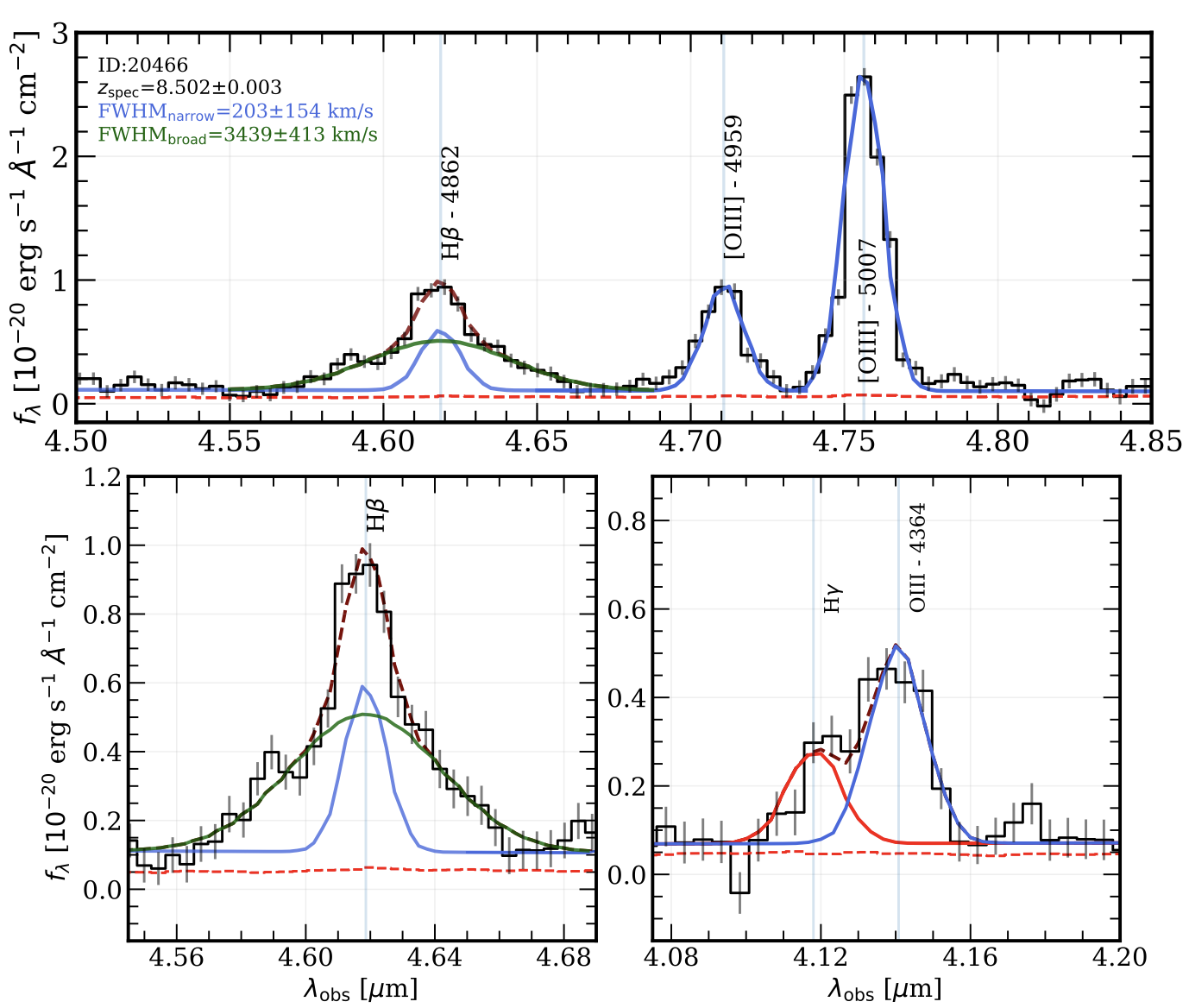}
\caption{\textbf{Top:} Best fit narrow (blue) and broad (green) line Gaussian components fit to the H$\beta$, [OIII] 4959,5007 line complex. A dual fit to the H$\beta$ is necessary to account for both the broad (FWHM$\sim3400$ km s$^{-1}$) and narrow (FWHM$\sim200$ km s$^{-1}$) components. The combined fit is shown as a dashed maroon line. The error spectrum is shown as a dashed red line.
\textbf{Bottom Left:} A zoom in on the H$\beta$ line.
\textbf{Bottom Right:} A fit to the blended H$\gamma$ (red) and auroral [OIII]$_{4364}$ (blue) lines. As before, the combined fit is shown as a dashed maroon line.}
\label{fig:fig2}
\end{center}
\end{figure*}

\subsection{Spectroscopic Redshift}
In order to compute the spectroscopic redshift, we use \textsc{msaexp} to fit sets of Gaussian continuum splines and emission lines templates to our spectrum.
As an input parameter, we set \texttt{nsplines=11} and allow to search for the minimal $\chi^2$ value across the a wide range of  $0.1\geq z \geq20$.
As indicated in \autoref{fig:fig1}, a clear Lyman break and several emission lines are clearly present in our spectrum, which results in an unambiguous redshift of $z_{\rm spec}=8.502\pm0.003$. This closely matches the initial redshift estimate of $z_{\rm phot}\sim9$ obtained by fitting the broad and medium band NIRCam data \citep{weaver23}
with \textsc{eazy} \citep{brammer08}.
\par
For the analysis presented in this work, we used an updated version (v1.1) of the \citet{furtak23} analytic strong lens model of the cluster. This new model includes one additional multiple image system in the northern substructure, and more importantly, an additional spectroscopic redshift in the north-western sub-structure from new VLT/MUSE observations of the cluster \citep{bergamini23}. Together with our secure $z_{\rm spec}$ for the source, we recover a magnification of $\mu=1.33\pm0.02$.

\subsection{Broad Line AGN Emission}
The spectrum of our object presented in \autoref{fig:fig1} shows a clear broad component present in the H$\beta$ emission line when contrasted to the narrower profiles of the adjacent [OIII] $\lambda \lambda$4959,5007 \AA\, lines. To evaluate the potential significance of this broad feature we perform a combined Gaussian fit to the entire line complex. For the two [OIII] lines we only consider a single narrow component, while the H$\beta$ is fit with both narrow and broad models. We allow the velocities of the narrow and broad components to vary between $50-500$ and $1000-5000$ km s$^{-1}$, respectively. In our fitting, we assume that the width of the narrow component for all three lines is the same, and a fixed ratio  [OIII]$_{5007}$/[OIII]$_{4959}=3$ \citep{storey2000}. In addition to that, we also model the local continuum with a first order polynomial. We fix the redshift to the one obtained from \textsc{msaexp}.
\par
We initialize the fit by first creating a set of models on the over-sampled wavelength grid. To mimic the variable resolution of the PRISM, we interpolate our model onto a variable step grid, while making sure that the total integrated flux is preserved. 
Early NIRSpec/MSA results \citep{de_graaff23} have shown that the spectral resolution of a point-like source falling within a slitlet is higher, compared to a uniformly illuminated slit, sometimes up to a factor of two. We therefore conservatively increase the nominal spectral resolution by a factor of 1.7. To take into account the effects of the line spread function, we additionally convolve our model with Gaussian of variable resolution \citep{de_graaff23,isobe23}.
The best fit is then found via non-linear least squares $\chi^2$ minimization. 
\par
From our fit we securely confirm the presence of a distinct broad component in H$\beta$, with a FWHM$=3439\pm413$ km s$^{-1}$. We measure the width of the narrow component to be $203\pm154$ km s$^{-1}$. In \autoref{fig:fig2} we show the best fit model to the entire line complex, with the narrow and broad components highlighted in blue and green, respectively. The flux of the broad component was measured to be $\sim 2$ higher than that of the narrow component.
\par
Emission lines in AGN are typically separated into permitted (e.g., Lyman and Balmer series of hydrogen) and (semi-) forbidden lines (e.g., [OIII]). The distinction between the two classes is thought to arise from two physically distinct regions around an actively accreting black hole (see \citealt{osterbrock78,vandenberk01}), the parsec scale broad-line region (BLR) and the kiloparsec-scale narrow-line region
(NLR). We find significant line broadening present in the $H\beta$ line, strongly hinting that some of the $H\beta$ emission originates from the BLR of a Type 1 AGN. While broad emission line features can also be an indicator of large-scale outflows, if such a scenario was indeed the case, a similar broadening would be present in the forbidden [OIII] lines \citep{amorin12,hogarth20}. We do not however find any evidence of that, on the contrary, the measured width of the narrow component is too low to be consistent with an outflow.

\subsection{Balmer Decrement}
The ratio between observed fluxes of Balmer series lines can be used to determine the dust extinction. For this purpose we will use the H$\beta$/H$\gamma$ ratio. As shown in the bottom right panel of \autoref{fig:fig1}, the H$\gamma$ line is blended with the [OIII]$_{4364}$. We however can use the prior information from the fit to the H$\beta$+[OIII]$_{4959,5007}$ complex and fix the narrow line widths to $\sim203$ km/s. Similarly to H$\beta$, the H$\gamma$ line would also contain a BLR component. However, the S/N of the broad component in H$\gamma$ is too low to reliably perform a double component fit. Due to this we assume that the primary contribution to the total H$\gamma$ flux is coming from the NLR, and compute the Balmer decrement by only considering ratio between 
the narrow component of the H$\beta$ line and our derived H$\gamma$ flux. In this case we ensure that both line fluxes are obtained by integrating similar velocity ranges. From this calculation we find H$\beta$/H$\gamma=3.2^{+0.7}_{-0.5}$.
\par
To compute the $A_{\rm V}$ we adopt the Small Magellanic Cloud (SMC) reddening law \citep{gordon03}, which has been found to match well the dust attenuation in high-$z$ galaxies \citep{capak15,reddy15,reddy18} and reddened quasars \citep[e.g. see][]{hopkins04}. Given Case B recombination, the intrinsic ratio between the lines is (H$\beta$/H$\gamma$)$_{\rm int}$=2.14 \citep{osterbrock89}. We find that the observed line ratio implies a high attenuation of $A_{\rm V}=2.1^{+1.1}_{-1.0}$.

\subsection{Template Fitting} \label{sec:template}
While it is likely that the contribution from the host galaxy to the total flux is small, as indicated by the mostly point-source like morphology of the source, obtaining an upper limit on the $M_*$ can still yield some crucial clues regarding the nature of our source. We follow the joint template fitting procedure described in \citet{labbe23}, fixing the redshift at $z_{\rm zspec}=8.502$. Briefly, the \citeauthor{labbe23} method employs a custom fitting procedure which combines dust-obscured stellar population models from \textsc{FSPS} \citep{conroy09} with empirical AGN models based on composite optical/near-infrared spectra of SDSS quasars \citep{vandenberk01,glikman06}. To fit the stellar population we utilize three independent FSPS models: a dust-free and dust obscured star forming components together with an old quiescent component. A constant SFH is assumed for the star forming components, while the quiescent part uses an exponentially declining SFH with $\tau=1$ Myr.
These templates are reddened by $A_{\rm   V}=0-5$ using a \citet{calzetti01} attenuation law and $R_V=3.1$. The attenuated AGN light is re-emitted in the mid-to-far infrared and modeled with the \textsc{CLUMPY} torus models \citep{nenkova08a,nenkova08b} as included with \textsc{FSPS}. The far-infrared dust emission associated with dust-obscured star formation is based in a set of \citet{draine07} templates employed within \textsc{FSPS}. NIRSpec observations are simulated using the published line spread function, increased by a factor 1.4 to account for a fact that this is a point source. For a full description of the fitting procedure, assumptions and potential caveats please refer to Appendix A in \citet{labbe23}.
\par
With these models we fit the NIRCam spectroscopy, all available NIRCam broad and medium band photometry, plus the point source flux from ALMA Band 6 at 1.2 mm \citep{fujimoto23,fujimoto23_alma_uncover} extracted using the NIRCam location as a prior. No flux is detected in ALMA to $(<70$ $\mu$Jy, $2\sigma)$, which strongly limits the contribution of massive star formation \citep[see][]{labbe13}. 
\par
Three types of model fits are performed: AGN-only, AGN+stars, stars-only. The AGN-only model fits a separate unreddened blue and reddened AGN template. Two templates are needed because of the remarkable dichotomy that the SEDs of typical ``litte red dots'' display, where the SED at $1-2$ $\mu$m (1000 - 1000 \AA\, rest) is blue ($f_\lambda\propto\lambda^{-2}$) while the SED at $3-5$ $\mu$m (3100 - 5200 \AA\, rest) is red 
 ($f_\lambda\propto\lambda^{0-2}$).
 \par
 The best-fit blue component is only $\sim1$\% of the bolometric luminosity of the red component, consistent with being scattered AGN light. The absence of clearly detectable broad UV lines (e.g., CIV, MgII) is consistent with expectations from the blue SDSS QSO template given the simulated NIRSpec PRISM resolution and depth. The best-fit dust attenuation is $A_{\rm V}\sim1.9$, when converted from \citet{calzetti01} to the SMC law with $R_V=2.7$, consistent with estimates based on the Balmer Decrement. Secondly, joint AGN+stars fit are performed. These should yield the most realistic constraint on the stellar mass from an underlying host galaxy. From these joint fits we derive an upper limit on the stellar mass of log$_{10}(M_*/M_\odot)<8.3$ based on the $95\%$ percentile of the posterior distribution of the mass of the stellar components. The stellar mass is primarily constrained by the combination of red NIRCam colors and the ALMA non-detection, which limits the amount of cold dust emission.
 We also note that evolved stellar populations are not seen in our spectrum, due to a lack of a Balmer break. As such the red continuum will most likely correspond to the dusty star formation, therefore allowing ALMA to further constrain the shape of our SED (see \citealt{labbe23} for further details). To derive an upper limit on the stellar mass, we assume that all emission originates from stars. For the stars-only fit we find log$_{10}(M_*/M_\odot)<8.7$ based on the $95\%$ of the posterior on the total stellar mass. We adopt this as a conservative upper limit on the $M_*$. 
 \par
As an additional test, we hypothesize what could be the maximum amount of the $M_*$ present in this object if we assume a more evolved stellar population and ignore the observed shape of the SED. For this we use a dust free model based on the exponentially declining SFH ($\tau=1$ Myr), that has formed 300 Myr ago at $z\sim15$. This is then scaled directly to the depth of our NIRCam images at the position of 20466 \citep{weaver23}. The resultant limit is then log$_{10}(M_*/M_\odot)<9.3$. We note however that such a model is not consistent with the photometry or the spectrum.
 \par
 Finally, it is also possible to envision that the AGN can be embedded within an extremely dusty galaxy, which surrounds the object and yet remains completely \textit{JWST}-dark, due to dust obscuration and low surface brightness. The implications of such dusty galaxies existing at high-$z$ have been briefly explored in \citet{kokorev23}, who present an overview of a highly obscured ($A_{\rm V}\sim4$) galaxy at $z=2.58$. By adapting the full UV to sub-mm SED from \citeauthor{kokorev23} and scaling it to our ALMA photometry, we compute a stellar mass limit of log$_{10}(M_*/M_\odot)<9.5$. This might be an overestimate, because the low redshift has a best-fit stellar age of 0.53 Gyr, implying that all mass is formed at $z=50$, which is not realistic. While definitive evidence for a significant presence of dust in galaxies only 500 Myr after the Big Bang is still lacking, deeper NIRCam observations of lensed fields could assist in uncovering the diffuse and dusty host galaxies surrounding reddened AGN.

\subsection{Other Emission Lines}
Beyond the lines discussed in earlier sections, our spectrum reveals emissions from Ly$\alpha$, the MgII doublet, and [NeIII] $\lambda \lambda$3869,3968. In this section, we outline the methods and underlying assumptions employed to determine the final flux of each line. All measured line fluxes can be found in \autoref{tab:tab2}.
\par
The Ly$\alpha$ is fit with a single Gaussian model, where the FWHM is allowed to vary. Using this fit we extract both the line intensity and the rest-frame equivalent width (EW$_0$), with the latter being measured at 240$\pm$30 \AA. 
\par
In our spectrum we identify the MgII $\lambda \lambda$2796,2804 doublet feature. It is expected that, similarly to H$\beta$, this permitted transition will contain both the narrow and broad components. However, the spectral resolution of PRISM at the observed wavelength of MgII does not allow us to securely separate the lines and perform a multiple component fit. We therefore fit the doublet with a single Gaussian model, where the FWHM is allowed to vary.
\par
The [NeIII] $\lambda \lambda$3869,3968 lines are nicely separated in our spectrum as seen in \autoref{fig:fig1}. As discussed previously, the emission from the forbidden lines is generally expected to arise from the NLR \citep{vandenberk01}, as such we fix the FWHM of [NeIII] to be the same as that of the [OIII] $\lambda \lambda$4959,5007.
\par
In addition to the detected lines, we also report a 3$\sigma$ upper limit for the low S/N ($<3$) [OII] $\lambda \lambda$3727,3729 doublet and H$\delta$.

\subsection{Size Measurement}
As can be seen in \autoref{fig:fig1}, the source is very compact and is just barely resolved in our imaging. We measure the effective radius in all available JWST bands 
using \textsc{GALFIT} \citep{peng02,peng10}, accounting for the effect of the empirically measured \citep{weaver23} point spread function. The light is modeled with a S\'ersic profile with the center, brightness, effective radius, S\'ersic index, and axis ratio as free parameters. From the best-fits we find that the measured sizes in short wavelength bands (F115W, F150W, F200W) are comparable to the ones at the long wavelength (F277W, F356W, F410M, F444W) and range from 0\farcs{035} to 0\farcs{04}, which is consistent with or even smaller than the FWHM of the NIRCam PSF for those bands. The consistency of sizes across all bands, coupled with a pure PSF fit being sufficient to describe the light from 20466 provides a further evidence of high degree of central concentration in our object. Finally, we derive the physical effective radius to be 165$\pm$20 pc, when corrected for the lensing magnification.


\begin{figure}
\begin{center}
\includegraphics[width=.45\textwidth]{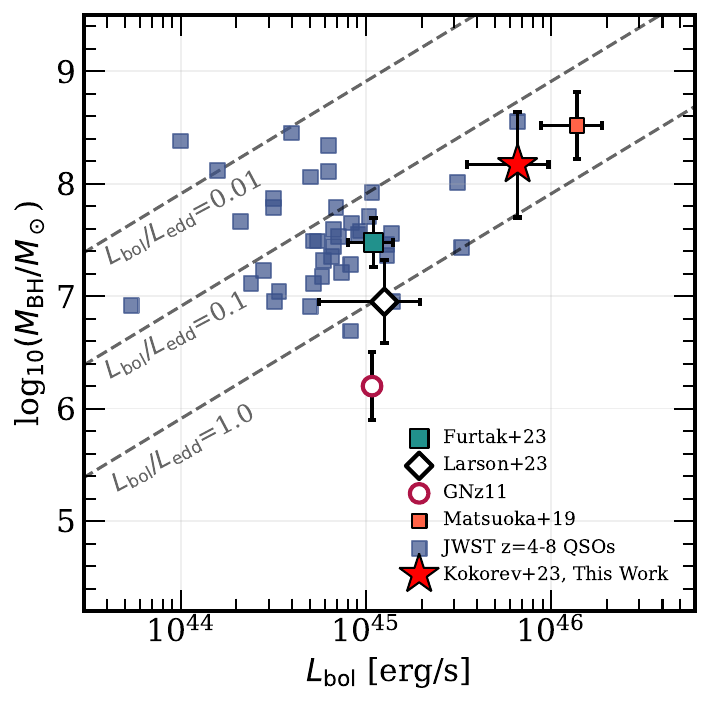}
\caption{Derived $M_{\rm BH}$ and $L_{\rm bol}$ compared to the other high-$z$ quasars. Our source is shown as a red star. Two high-$z$ AGN candidates, CEERS\_1019 at $z=8.7$ \citep{larson23} and GNz11 at $z=10.6$ \citep{maiolino23} are shown as open diamond and circle, respectively. JWST detected AGN at
$z>4$ are shown as blue squares \citep{kocevski23,harikane23_agn,maiolino23b,matthee23}. The cyan square shows the triply lensed quasar from \citet{furtak23_nat}. A massive, bright $z=7.07$ QSO from \citet{matsuoka19} is shown as an orange square. The dashed lines show the bolometric luminosities with
the Eddington ratios of $L_{\rm bol}/L_{\rm edd}$ = 0.01, 0.1 and 1.0.}
\label{fig:fig3}
\end{center}
\end{figure}

\begin{deluxetable}{ccc}[]
\tabcolsep=.4mm
\tablecaption{\label{tab:tab2}
Measured Line Fluxes}.
\tablehead{Line & $\lambda_{\rm rest}$ [\AA] & Flux [10$^{-20}$ erg s$^{-1}$ cm$^{-2}$]}
\startdata
Ly$\alpha$ & 1215.4 & $414.1\pm85.2$ \\
MgII & 2796.5,2803.1 & $62.5\pm15.1$ \\
$[$OII$]$ & 3727.0,3729.9 & $15.2\pm8.0$ \\
$[$NeIII$]$ & 3869.9 & $62.2\pm15.4$ \\
$[$NeIII$]$ & 3968.6 & $38.9\pm9.2$ \\
H$\delta$ & 4102.9 & $12.2\pm6.5$ \\
H$\gamma$ (narrow+broad) & 4341.7 & $24.6\pm3.1$ \\
$[$OIII$]$ & 4364.4 & $83.0\pm7.2$ \\
H$\beta$ (narrow) & 4862.7 & $78.6\pm5.9$\\
H$\beta$ (broad) & 4862.7 & $232.6\pm17.3$\\
$[$OIII$]$ & 4959.5 & $136.7\pm3.0$\\
$[$OIII$]$ & 5007.2 & $412.6\pm11.3$\\
\enddata
\end{deluxetable}

\begin{figure}
\begin{center}
\includegraphics[width=.45\textwidth]{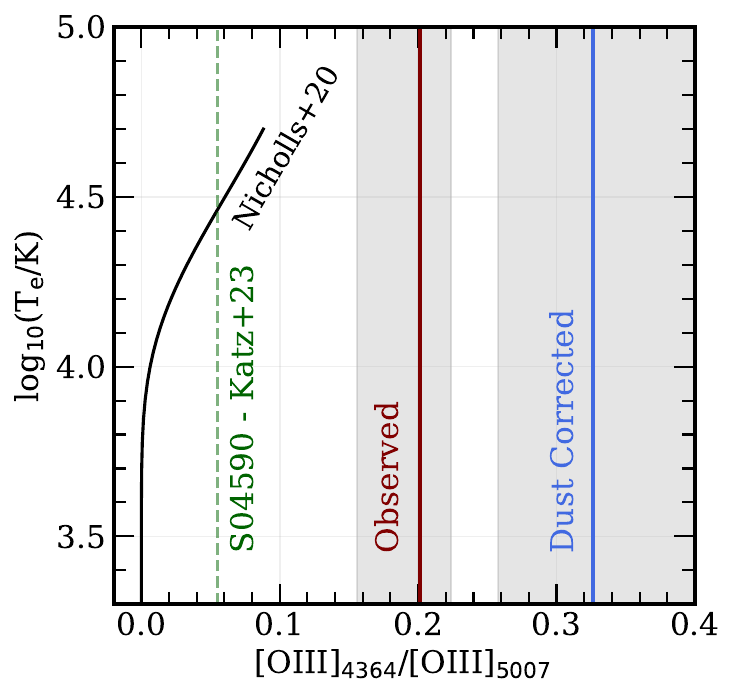}
\caption{The electron temperature 
(T$_{\rm e}$) as a function of [OIII]$_{4364}$/[OIII]$_{5007}$ (RO3) ratio. In black we present models from \citet{nicholls20} for different electron densities ranging from 1 to $10^4$, the dynamic range does not allow to distinguish between different lines however. We show our observed line ratio in maroon, and in blue after the dust correction. The $z=8.5$ source displaying extreme RO3 from \citet{katz23} is shown with a dashed green line. The shaded envelopes show the 1$\sigma$ uncertainty. The extremely high line ratio we find, is indicative of AGN dominated ionization.}
\label{fig:fig_te_auroral}
\end{center}
\end{figure}

\begin{figure}
\begin{center}
\includegraphics[width=.45\textwidth]{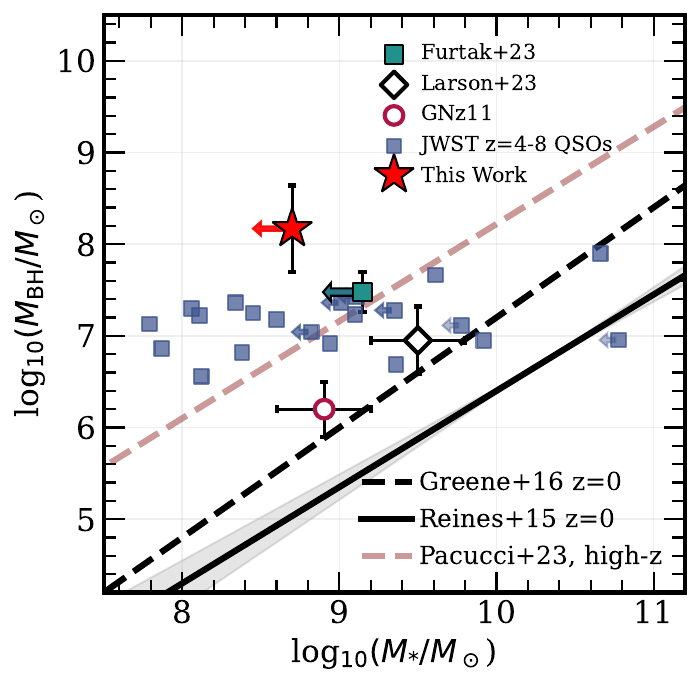}
\caption{Black hole to stellar mass relation. The color coding is the same as in \autoref{fig:fig3}. The solid and dashed black lines show best fits to the $z=0$ AGN samples from \citet{reines15}
and \citet{greene16}, respectively. The high-$z$ trend from \citet{pacucci23} is shown as a dashed maroon line.  Only an upper limit on $M_*$ is available for our source and the triply lensed quasar from \citet{furtak23_nat}.}
\label{fig:fig_mbh_m*}
\end{center}
\end{figure}

\begin{figure*}
\begin{center}
\includegraphics[width=.95\textwidth]{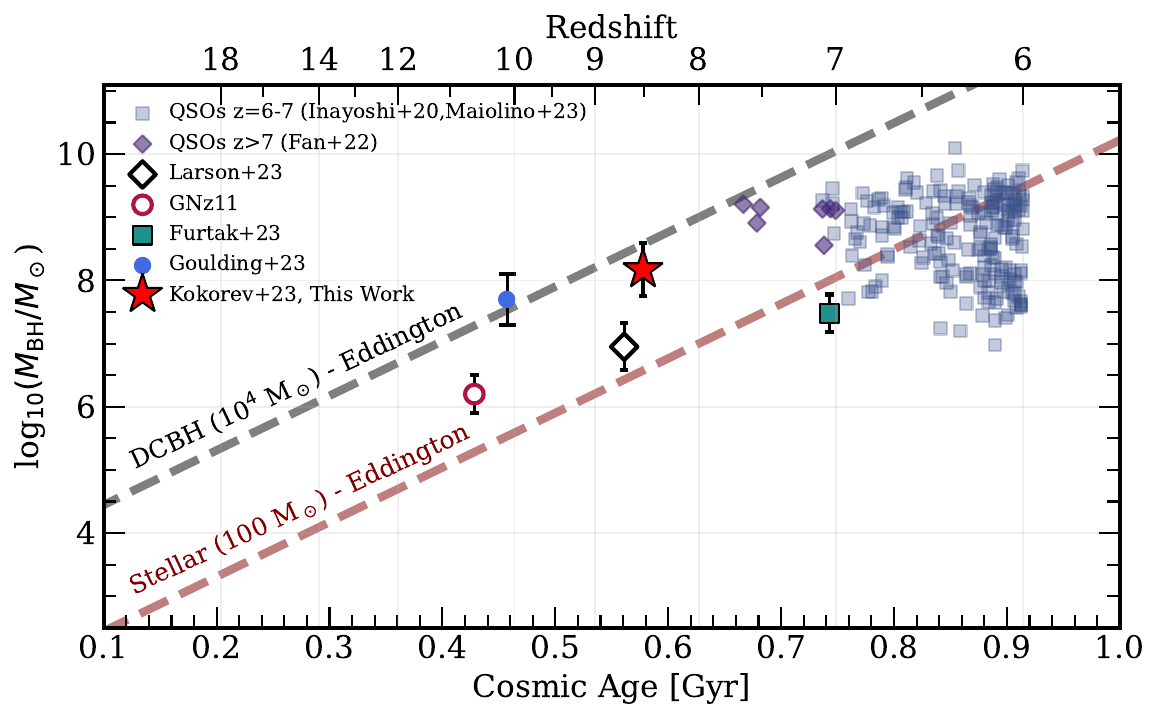}
\caption{Black hole mass versus cosmic age. The red star shows the result presented in this work. The dark blue squares and violet diamonds show the results for $z=6-7$ and $z>7$ QSOs respectively, as presented in 
\citet{inayoshi20,maiolino23b} and \citet{fan22}. Triply lensed $z=7$ QSO from \citet{furtak23_nat} is shown as a cyan square. A tentative $M_{\rm BH}$ range for UHZ1 at $z=10.1$ from \citet{goulding23} is shown as a blue circle. Two high-$z$ AGN candidates, namely $z=8.7$ AGN from CEERS \citep{larson23} and GNz11 at $z=10.6$ \citep{maiolino23} are shown as open diamond and circle, respectively. Dashed lines show analytical models of blackhole mass accretion at an Eddington rate starting from a stellar seed (maroon) and a direct collapse (DCBH) model (black).}
\label{fig:mbh_z}
\end{center}
\end{figure*}

\section{Results}
\label{sec:results}
\subsection{AGN Properties}
Reverberation mapping of quasars has revealed a correlation between the distance to the BLR in AGN and the width of the H$\beta$ line \citep[e.g.][]{kaspi00,greene05}, which allows for the black hole mass to be estimated from single-epoch measurements. Using the standard relations presented in \citet{greene05} we compute the black-hole mass ($M_{\rm BH}$) from both the luminosity and the width of the broad component of the H$\beta$ line and the rest-frame $L_{5100}$ luminosity of the continuum. 
\par
Using our best fit models for lines and the continuum and accounting for the effects of lensing and dust obscuration, we find log$_{10}(M_{\rm BH, H\beta}/M_{\odot})=8.17\pm0.42$ and log$_{10}(M_{\rm BH, 5100}/M_{\odot})=8.01\pm0.40$. The uncertainties in this case are primarily dominated by the scatter ($\sim 0.3$ dex) in the virial relation used to derive the mass, rather than the errors on the measurements themselves (e.g. see \citealt{kollmeier06}). From the fact that both methods return a consistent result we can conclude that the rest-frame 5100 \AA\, luminosity of 20466 is almost entirely dominated by the light from the AGN. This allows us to further justify the assumptions made when modeling the spectrum in \autoref{sec:template}. We adopt the  $M_{\rm BH}$ derived from H$\beta$ line luminosity as our final result.
\par
In addition to the mass, the broad H$\beta$ can be used to compute the bolometric luminosity ($L_{\rm bol}$) of the AGN. Typically $L_{\rm bol}$ is computed from the luminosity of H$\alpha$, which we lack.  However we can assume a broad-line intrinsic ratio of H$\alpha$/H$\beta\sim3.06$ \citep{dong08}, and use it in conjunction with $L_{\rm bol}=3.06\times 130\times L_{H\beta}$ \citep{richards06,stern12}. Assuming that the dust attenuation within NLR and BLR are similar, we correct the H$\beta$ luminosity for dust attenuation and magnification effects, thus obtaining $L_{\rm bol}=(6.6\pm3.1)\times10^{45}$ erg s$^{-1}$. These results suggest that our object is accreting at $\sim 40$ \% of the Eddington rate, with $L_{\rm bol}/L_{\rm edd}\sim0.4$, suggesting a sub-Eddington accretion scenario. All of the estimated parameters are presented in \autoref{tab:tab1}.
\par
In \autoref{fig:fig3} we show the estimated $M_{\rm BH}$ and $L_{\rm bol}$ for our object. The mass exceeding $10^8$ $M_\odot$, and luminosity brighter than $5\times10^{45}$ erg s$^{-1}$, indicate that our source is more massive and luminous than the majority of quasars at $z\sim 4-8$ range identified with JWST \citep{furtak23_nat,harikane23_agn,kocevski23}. The object presented in \citet{larson23} at $z\sim8.7$ is only $\sim20$ Myr younger, however shows an almost 1 dex difference in the derived BH mass. This suggests that CEERS\_1019 and 20466 might have followed vastly different evolutionary scenarios.
On the other hand, we find similar (within 1$\sigma$) mass and luminosity when compared to a $z=7.07$ quasar presented in \citet{matsuoka19}.

\subsection{Ionization Mechanisms}
In this section we briefly explore the ratios between the measured emission lines to investigate the potential ionization mechanisms in our source.
We observe an unusually high ratio between the auroral [OIII]$_{4364}$ line and [OIII]$_{5007}$ (RO3) of 0.32. In \autoref{fig:fig_te_auroral} we explore the typical ``allowed" RO3 for a range of electron temperatures $T_{\rm e}$ and densities $n_{\rm e}$ from the models presented in \citet{nicholls20}, alongside our observed and dust corrected  RO3. When compared to the models, our object appears to be a significant outlier, regardless of the adopted $T_{\rm e}$ and $n_{\rm e}$ values. 
Extreme values of RO3 have already been reported in recent JWST spectra, for example a $z=8.5$ galaxy presented in \citet{katz23} shows RO3 of 0.048, when corrected for dust. Elevated RO3 are not new and have been discussed in the context of low-$z$ Seyfert galaxies \citep[e.g.][]{koski76,osterbrock78,ferland83,dopita95,nagao01,baskin05,binette22}. In fact, the photoionization models of \citet{baskin05} suggest that it is possible to reach the required densities and temperatures to produce extreme RO3 within the NLR around an AGN.
\par
In addition we can investigate our source in the context of the often utilized ``OHNO" diagnostic, which compares the [OIII]$_{5007}$/H$\beta$ and [NeIII]$_{3869}$/[OII]$_{3727,3729}$ ratios. This diagnostic has been used at low and high-$z$ to ascertain whether the ionization is powered purely by star formation  or by an AGN \citep{backhaus22,cleri22,larson23}. After dust correction, we find log$_{10}$([OIII]$_{\rm 5007}$/H$\beta$)$\sim0.68$ and a lower limit of log$_{10}$([NeIII]$_{\rm 3870}$/[OII])$>0.15$. These line ratios, while not as high as reported in other $z>8$ AGN candidates \citep{larson23}, are still indicative of high ionization in 20466.

\subsection{Massive Accreting Black Hole at $z=8.50$}
With all the information in hand, we would like to remark on the most probable nature of our source. The spectrum of 20466 shows a clear broad line component present in the $H\beta$ line. With a confidence level of $13.4\sigma$ we estimate the FWHM of the broad profile to be $\sim 3440$ km s$^{-1}$. The adjacent [OIII] lines are well fit with a much narrower Gaussian ($\sim 200$ km s$^{-1}$), ruling out potential outflows (e.g. see \citealt{chisholm15}. This leads us to conclude that broad H$\beta$ emission most likely originates within a BLR of an AGN. Furthermore, we note high ionization present in our source, as indicated by the extreme RO3 ratio as well as the ``OHNO" diagnostic. These values are difficult to reconcile with photoionization by young stars alone, highlighting the strong ionizing nature of this source. Combining the above with a near point-like source morphology observed across all NIRCam bands leads us to conclude that 20466 is a massive and luminous Type - 1 AGN, observed just 580 Myr after the Big Bang.

\section{Discussion and Conclusion}
\label{sec:conclusion}

\subsection{Black Hole Formation Mechanisms}
A significant unresolved question which remains to be answered, is how these supermassive black holes come to be. In \autoref{fig:fig_mbh_m*} we present the $M_{\rm BH}$ vs an upper limit on the $M_*$, derived from the template fitting, compared to the relations for QSOs at $z\sim0$, JWST detected AGN at $z=4-5$, as well as those recently identified at high-$z$. We find an extremely high ratio of black hole to host mass of at least $\sim 30 \%$, which is orders of magnitude higher compared to the local QSOs, and also elevated when compared to other massive AGN at high-$z$. Predictions from ``direct collapse black hole'' (DCBH) formation models suggest that this ratio can indeed be high close to the seeding epoch \citep{natarajan11,natarajan17}, compared to the local values. Moreover, reaching high $M_{\rm BH}/M_*$ from light seeds would require growing the $M_{\rm BH}$ without also growing the galaxy mass, which does not seem feasible. 
This gives us a first hint regarding the potential formation and accretion mechanisms of our source.
\par
We further explore this in \autoref{fig:mbh_z} which presents the $M_{\rm BH}$ as a function of cosmic age for our AGN, alongside previous measurements for high-$z$ QSOs. It is possible to explain the existence of these SMBH via the super-Eddington accretion rates \citep[e.g.][]{haiman01,alexander14} in low mass black hole seeds ($\sim100$ $M_\odot$), formed from the Population III stars \citep{madau01}. Another possible solution is to start from heavy DCBH seeds $\sim10^{3-5}$ $M_\odot$ from the collapse of pristine gas in early halos \citep[e.g.][]{rees84,loeb94}.
In this work we consider both options, starting with low (100 $M_\odot$) and high ($10^{4}$ $M_\odot$) mass seeds at $z\sim50$, we explore different accretion rates required to grow a black hole to 10$^{8}$ $M_\odot$ by $z=8.5$.
In \autoref{fig:mbh_z} we show that the observed $M_{\rm BH}$ can be reproduced by continuous Eddington accretion driven growth in a DCBH seed at 10$^{4}$ $M_\odot$, with the Eddington limited stellar mass seed scenario being unlikely. Furthermore, it would be difficult to explain a radical change in $M_{\rm BH}/M_*$ over these early times, which would naively mean that the high black hole to galaxy ratio that we find also favors heavy seeds.
Alternatively, simulations show that it is also possible to grow an SMBH via super-Eddington accretion onto a stellar mass seed \citep{jeon12,massonneau23}, however the feasibility of such extreme accretion modes is yet to be conclusively determined. 
\par
Recent detection and NIRSpec observations of an X-ray quasar at $z=10.07$ - UHZ-1 \citep{bogdan23,goulding23} find a $M_{\rm BH}/M_*$ $\sim5-100 \%$, comparable to the ratio we derive in this work. MSA ID 20466 and UHZ-1 are located in the same field, in a relatively small area,  which suggests that black holes that form from direct collapse could be more common than previously thought and some early overmassive SMBHs may indeed originate from heavy seeds \citep{eilers23,natarajan23,pacucci23,stone23}.
\par
However, we note that the elevated $M_{\rm BH}$ to host stellar mass trends, might not manifest when using the dynamical mass of the galaxy instead. A recent work by \citet{maiolino23b} finds little deviation between $M_{\rm BH}$-$M_{\rm dyn}$ relation in $z>4$ broad line AGN and that of local QSOs. This in return might imply that the black hole formation is more strongly connected with the mass assembly itself, rather than star formation in the host galaxy.

\subsection{Final Remarks}

Using the NIRSpec/PRISM and NIRCam data from JWST UNCOVER survey we present the discovery of an actively accreting SMBH at $z=8.502\pm0.003$. The spectrum of our object shows an unambiguous ($>10\sigma$) broad-line component  present in the H$\beta$ line, exhibiting a FWHM of $\sim3400$ km s$^{-1}$. Although comparable velocities could potentially stem from large scale outflows, a corresponding broadening effect should manifest in the adjacent [OIII]$_{4959}$ and [OIII]$_{5007}$ lines. We do not, however, find any evidence which can support this. By examining the RO3 and the ``OHNO" line diagnostic, we find values which are consistent with high ionization present in the source. These findings, compounded by near point source morphology, lead us to deduce that the underlying cause behind the extended broad-line region can only be attributed to the AGN activity.
\par
From the flux and FWHM of the H$\beta$ we compute the black hole mass of log$_{10}(M_{\rm BH}/M_\odot)=8.17\pm0.42$ and a luminosity of $L_{\rm bol}=6.6\pm3.1$ erg s$^{-1}$, suggesting 
an accretion rate at $\sim 40\%$ of the Eddington limit. We also find that the $M_{\rm BH}$ derived from the H$\beta$ is consistent to the one computed from rest-frame 5100 \AA\ continuum within one standard deviation, potentially indicating that the spectrum of this object is strongly, or almost entirely dominated by the AGN emission.
\par
Finally, we explore multiple scenarios which can lead to a presence of such massive BH by $z$=8.50. We find that the BH mass can not be reproduced by Eddington-limited accretion from a stellar seed, unless super-Eddington regimes can be achieved. It is worth noting that this would require a gas rich disk providing material for accretion while remaining unaffected by BH feedback. Even in our models that start at $z=50$, this SMBH would need to grow at sustained rates for $>400$ Myr, all the while remaining in a super-Eddington regime, which is 
beyond physically plausible scenarios. On the other hand, such a mass can also be attained via Eddington accretion driven growth for DCBH ($\sim10^{4}$ $M_\odot$) seeds. A direct collapse model would then also be able to potentially explain the extreme BH to host ratio of least $\sim 30$ \% found in our source.
\par
While their formation pathways are still largely uncertain, the high number of AGN detected with JWST at high-$z$, given the relatively small areas covered so far, implies that extremely massive SMBH are already in place at $z>7$, just $\sim700$ Myr after the Big Bang \citep{wang21,furtak23_nat,goulding23,greene23,larson23,pacucci23}. The clues of high black hole to stellar mass ratios, exceeding several tens of percent, will allow us to place stronger constraints on the sizes of BH seeds and accretion modes required to produce such massive objects \citep{bogdan23, goulding23}.
Interestingly, the Ly$\alpha$ line is clearly detected from 20466 despite its heavily dusty nature, and 20466 turns out to reside in a huge ionized bubble with a radius of 7.6 proper Mpc \citep{fujimoto23_uncover}.  
These results may indicate that the recent abundant AGN population identified with JWST provides a non-negligible contribution of ionizing flux to cosmic reionization. 
It is also possible that these AGN are residing in galaxy overdensities, which may have reionized the Universe earlier in their environments.
The identification of 20466, along with similar massive quasars at high-$z$, suggests we still lack a comprehensive understanding of AGN and host galaxy co-evolution in the early Universe, which we only recently started exploring with JWST.

\acknowledgments
We thank the anonymous referee for a number of
constructive suggestions, which helped to improve this manuscript. We thank Darach Watson and Maxime Trebitsch for insightful discussions on black hole physics. We are also grateful to Katriona Gould for helpful discussions about \textsc{msaexp}. We would like to thank Alex Pope for useful discussion regarding dusty star formation. VK and KIC acknowledge funding from the Dutch Research Council (NWO) through the award of the Vici Grant VI.C.212.036. AZ acknowledges support by Grant No. 2020750 from the United States-Israel Binational Science Foundation (BSF) and Grant No. 2109066 from the United States National Science Foundation (NSF), and by the Ministry of Science \& Technology, Israel. IL acknowledges support by the Australian Research Council through Future Fellowship FT220100798.

PD acknowledges support from the Dutch Research Council (NWO) through the award of the VIDI Grant 016.VIDI.189.162 (``ODIN") and the European Commission's and University of Groningen's CO-FUND Rosalind Franklin program. H.A. and IC acknowledge support from CNES, focused on the JWST mission, and the Programme National Cosmology and Galaxies (PNCG) of CNRS/INSU with INP and IN2P3, co-funded by CEA and CNES. RP and DM acknowledge support from JWST-GO-02561.013-A. YF acknowledge support from NAOJ ALMA Scientific Research Grant number 2020-16B. YF further acknowledges support from support from JSPS KAKENHI Grant Number JP23K13149. MS acknowledges support from the CIDEGENT/2021/059 grant, from project PID2019-109592GB-I00/AEI/10.13039/501100011033 from the Spanish Ministerio de Ciencia e Innovaci\'on - Agencia Estatal de Investigaci\'on. MST also acknowledges the financial support from the MCIN with funding from the European Union NextGenerationEU and Generalitat Valenciana in the call Programa de Planes Complementarios de I+D+i (PRTR 2022) Project (VAL-JPAS), reference ASFAE/2022/025. This work is based on observations made with the NASA/ESA/CSA James Webb Space Telescope. The data were obtained from the Mikulski Archive for Space Telescopes at the Space Telescope Science Institute, which is operated by the Association of Universities for Research in Astronomy, Inc., under NASA contract NAS 5-03127 for JWST. These observations are associated with program JWST-GO-2561. Support for program JWST-GO-2561 was provided by NASA through a grant from the Space Telescope Science Institute, which is operated by the Association of Universities for Research in Astronomy, Inc., under NASA contract NAS 5-03127. This work has received funding from the Swiss State Secretariat for Education, Research and Innovation (SERI) under contract number MB22.00072, as well as from the Swiss National Science Foundation (SNSF) through project grant 200020\_207349. The Cosmic Dawn Center (DAWN) is funded by the Danish National Research Foundation under grant No.\ 140. The work of CCW is supported by NOIRLab, which is managed by the Association of Universities for Research in Astronomy (AURA) under a cooperative agreement with the National Science Foundation. 

\software{EAZY \citep{brammer08}, FSPS \citep{conroy09}, GALFIT \citep{peng02,peng10}, grizli \citep{grizli}, msaexp \citep{msaexp}.}

\facilities{\jwst, ALMA}
\bibliographystyle{aasjournal}
\bibliography{refs}

\end{document}